# TITAN MAGNETIC TAIL: DOES ITS CONFIGURATION CORRESPOND TO THE INDUCED MAGNETOSPHERE?

P. Israelevich, and A. Ershkovich

Department of Geophysics and Planetary Sciences, Tel Aviv University, Tel Aviv, 69978, Israel.

P. Israelevich, and A. Ershkovich

P. Israelevich, Department of Geophysics and Planetary Sciences, Tel Aviv University, Tel Aviv, 69978, Israel. (peter@post.tau.ac.il)

A. Ershkovich, Department of Geophysics and Planetary Sciences, Tel Aviv University, Tel Aviv, 69978, Israel.

**Abstract.** Magnetic field structure observed in numerous *Cassini* flybys in the region of Titan interaction with the corotating flow of Kronian magnetosheric plasma contradicts the classical picture of the ideal induced magnetosphere produced by the magnetic field line draping about the obstacle. Clear draping is observed only upstream the Titan, but not in the Titan magnetic wake.

We consider the magnetic field tension downstream the Titan magnetic tail and show that the magnetic field direction is not consistent with the induced magnetosphere produced by magnetic field lines draping. We arrive at the conclusion that the mechanisms alternative to the induced magnetosphere formation should be considered for the Titan magnetic surrounding.

# 1. Introduction

Induced magnetosphere arises when a magnetized plasma flow interacts with a non-magnetic body possessing ionosphere. Electric currents are induced under the action of Lorentz electric field $\mathbf{E} = -\mathbf{v} \times \mathbf{B}$. Superposition of the external magnetic field and field produced by induced currents results in a configuration proposed by *Alfvén* (1957) for cometary tail and can be described be magnetic field lines draping around the obstacle. Magnetic field configuration of the magnetic tail in the classical induced magnetosphere is shown in Figure 1. It consists of two lobes of oppositely directed magnetic field lines separated by the current sheet. The upper lobe of the tail is shown in the figure.

Orientation of such an induced magnetosphere depends on the direction of the magnetic field in the plasma flow. If we consider the coordinate system $X_M Y_M Z_M$ with the unit vectors defined as

$$\begin{aligned} \mathbf{e}_z &= \mathbf{v}_0 / v_0 \\ \mathbf{e}_y &= (\mathbf{v}_0 \times \mathbf{B}_0) / |\mathbf{v}_0 \times \mathbf{B}_0| \\ \mathbf{e}_x &= \mathbf{e}_y \times \mathbf{e}_z \end{aligned} \qquad (1)$$

where $\mathbf{v}_0$ and $\mathbf{B}_0$ are flow velocity and magnetic field in the free flow upstream the obstacle, respectively. Simple criterion for ideal induced magnetosphere (field line draping) can be formulated as

$$\begin{aligned} B_z &> 0 \text{ for } x > 0 \\ B_z &< 0 \text{ for } x < 0 \end{aligned} \qquad (2)$$

and $B_y > 0$ inside the magnetosphere.

Moreover, the direction of projection of the magnetic field onto *XY*-plane is close to the *Y*-axis. Figure 2 shows the magnetic field vector projections onto induced magnetic tail cross section as obtained in single fluid MHD simulations (*Kabin et al.*, 2000). The clock angle of the magnetic field inside the induced magnetosphere does not exceed ~ 15$^o$ Such a configuration was observed in typical induced magnetosphere of Venus (*Dolginov et al.,* 1981; *McComas et al.*, 1986) and comet Halley (*Israelevich et al.*, 1994).

Titan orbit is located mostly inside the Kronian magnetosphere. Therefore Titan interacts with subsonic flow of corotating plasma (with rather rare exceptions when the satellite leaves the planetary magnetosphere and appears in the shocked solar wind flow). It is commonly believed that the induced magnetosphere results from interaction of Titan atmosphere with the stream of Kronian plasma. The magnetic field perturbation in the Titan wake was first observed by Voyager-1 (*Ness et al.,* 1982) (Figure 3). However, Voyager observations do not satisfy the above criteria of an ideal induced magnetosphere. Strong rotation of the magnetic field around the flow direction (*Z*-axis) occurred in the Titan magnetic tail and the direction of flow aligned component of the magnetic field does not correspond to the field line draping (*Kabin* et al., 2000). *Kabin* et al., (2000) ascribed this violation of draping criterion to possible influence of Titan's own magnetic field.

In this paper we will show that field structure in the Titan magnetic tail in most cases contradicts the filed line draping picture. We will also apply a new approach which does not require the knowledge of the magnetic field direction in the free flow in order to judge whether or not the field topology corresponds to draping.

**2. Magnetic field configuration in Titan wake**

Systematic study of numerous Cassini flybys also does not show clear picture of the induced magnetosphere. *Simon et al.* (2013) divided the flybys into groups according to their agreement with the draping criterion (2). Positions of closest approaches for different flybys are shown in Figure 4 (circles are used for the flybys with small perturbations in the corotating flow, crosses - for the cases of strong perturbations. In the first case checking draping criterion was rather reliable, in the second case - somewhat dubious. Blue colour corresponds to the flybys where draping geometry was observed, red colour - when the draping criterion was violated).

Ideal draping pattern was observed (with one exception) for the flybys with closest approaches on the upstream ($z < 0$) side of Titan, and not in the tail.

The verification of draping criterion (2) is intricate because the magnetic field direction in the free flow is not known for the time periods when *Cassini* was inside the Titan magnetosphere, so the data cannot be presented in true coordinate system $X_M Y_M Z_M$ defined by (1). However, if one assumes that the clock angle of the magnetic field in the wake does not vary significantly, the direction of the local magnetic field may be used for an approximate determination of the $Y_M$-axis direction, and the system with unit vectors

$$\begin{aligned} \mathbf{e}_z &= \mathbf{v}_0 / v_0 \\ \mathbf{e}_y &= (\mathbf{v}_0 \times \mathbf{B}) / |\mathbf{v}_0 \times \mathbf{B}| \\ \mathbf{e}_x &= \mathbf{e}_y \times \mathbf{e}_z \end{aligned} \qquad (3)$$

can be used as a proxy for $X_M Y_M Z_M$-system. (Here, **B** is the local measured magnetic field).

Data for *Cassini* flybys have been converted into this proxy coordinate system. Figure 5 shows the dependence of the $B_z$ magnetic field component on the satellite $Y_M$ coordinate. Top panel, showing this dependence on the upstream side of Titan ($Z < 0$) reasonably well corresponds to the draping criterion (2). Data for the downstream side are presented in the bottom panel. Obviously, the draping criterion is violated on the downstream side ($Z > 0$).

**3. Magnetic field tension in the Titan tail**

However, the criterion (2) may be obscured in the real induced magnetosphere by variations of corotation flow and upstream magnetic field directions which are not known for the period of time when the spacecraft was inside the Titan magnetosphere. Here we will apply another test for induced nature of the magnetic field in the Titan tail which is not subjected to flow/upstream magnetic field variability.

As can be seen from Figure 1, centers of curvature of the field line forming the induced magnetosphere is always located downstream of the line. Thus, magnetic field tension $(\mathbf{B}\nabla)\mathbf{B}$ is always directed downstream in the induced tail produced by the field line draping and hence it is an inherent feature of the ideal induced magnetosphere. Thus, it may be used in order establish the nature of the magnetosphere as induced by Lorentz electric field. Application of this method allows us to avoid the influence of magnetospheric changes due to upstream magnetic field variations.

We estimate the value of the magnetic tension density as $(\mathbf{B}\nabla)\mathbf{B} \approx B_x \frac{\delta B_z}{\delta x} + B_y \frac{\delta B_z}{\delta y}$ along the Cassini trajectories. Figure 6 shows the direction of the tension force inside the geometrical shadow of Titan downstream the satellite. Blue points correspond to the tension force along the flow, as it is expected for the induced magnetosphere, and red points - to the tension against the flow, contradicting the magnetic field line draping. Distribution of the tension direction is rather chaotical, but in average the magnetic tension is directed against the flow rather than downstream.

Figure 7 shows the histogram of distribution of the magnetic tension. Negative values correspond to the tension against the flow which contradicts the draping nature of the Titan magnetic tail. The mean value of the tension is ~ 0.5 $nT^2/R_T$ against the plasma flow. This pattern does not correspond to the picture of the induced magnetosphere.

## 4. Conclusions

We arrived at the conclusion that *Cassini* magnetic field measurements during Titan flybys do not correspond to the ideal picture of the induced magnetosphere. This discrepancy requires taking into consideratoin alternative or additional mechanisms responsible for the Titan interaction with Kronian plasma corotation flow. They may include: (1) intrinsic (crustal) magnetic field of Titan, (2) significance of Hall effect in the process of interaction, (3) influence of atmospheric neutral winds on the magnetospheric structure, and (4) effects of electromagnetic induction due temporal variations of the magnetic field ($\partial\mathbf{B}/\partial t$) in the magnetospheric plasma flow.

**List of Figures**

Figure 1. Configuration of an ideal induced magnetosphere resulting from the magnetic field line draping about the conducting obstacle. The magnetic tail consists of two lobes with oppositely directed magnetic field. The field lines are shown only above the current sheet.

Figure 2. Projections of magnetic field vectors in the wake of the induced magnetosphere onto the plane perpendicular to the stream velocity: (a) at the distance of $1R$ and (b) $1.85R$ behind the obstacle of radius $R$.

Figure 3. Plots of magnetic field vector components along the trajectory of *Voyager* -1 (*Ness et al.*, 1982)

Figure 4. Positions of the Cassini closest approaches in Titan-centric coordinates. Circles (crosses) denote cases with small (moderate) perturbations in the corotating flow. Blue color corresponds to draping geometry and red color - to the violation of draping criterion.

Figure 5. : Dependence of the flow aligned magnetic field component $B_z$ on the distance along the $Y_M$ axis. (Top) for the measurements upstream from the obstacle, (bottom) for downstream measurements. Blue dots - individual measurements, red dots - averages for $0.2\ R_T$ bins along the $Y_M$ coordinate.

Figure 6. Direction of magnetic tension force in the Titan tail. (Blue dots along the flow, red dots against the flow, correspondingly).

Figure 7. Histogram of distribution of magnetic tension in the region of Titan wake. Only the points at the distances less than 1 $R_T$ from the tail axis are taken into account.

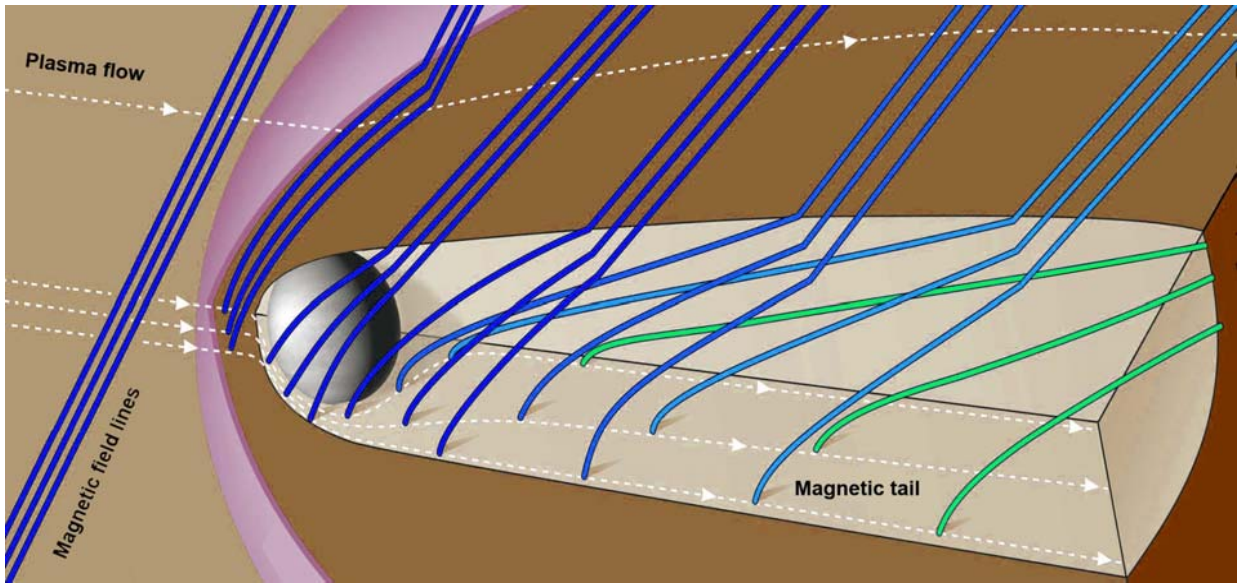

Figure 1: Configuration of an ideal induced magnetosphere resulting from the magnetic field line draping about the conducting obstacle. The magnetic tail consists of two lobes with oppositely directed magnetic field. The field lines are shown only above the current sheet.

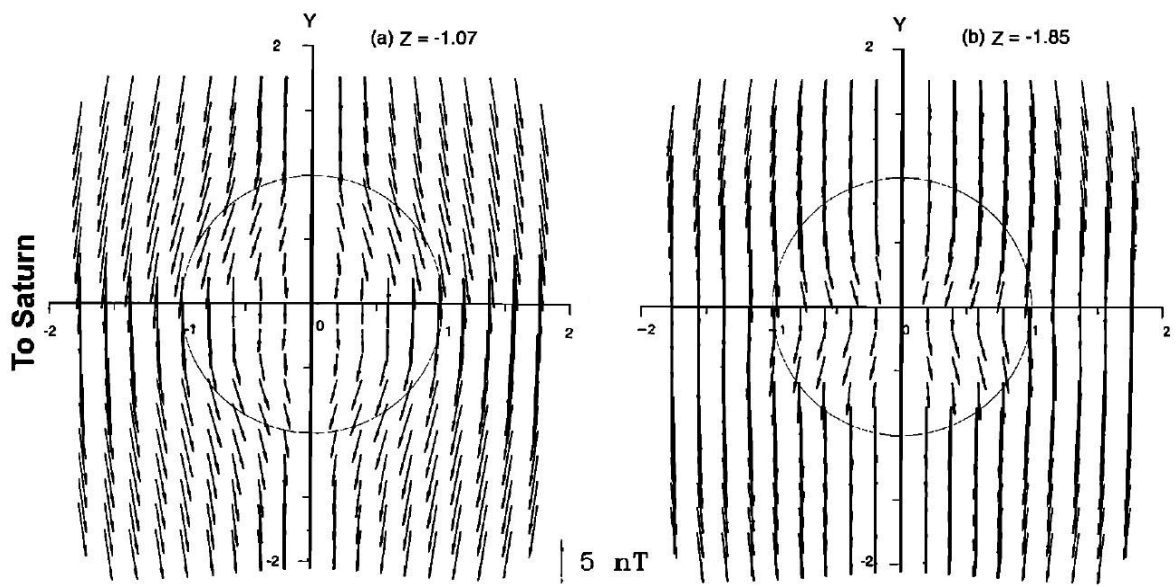

Figure 2: Projections of magnetic field vectors in the wake of the induced magnetosphere onto the plane perpendicular to the stream velocity: (a) at the distance of $1R$ and (b) $1.85R$ behind the obstacle of radius $R$.

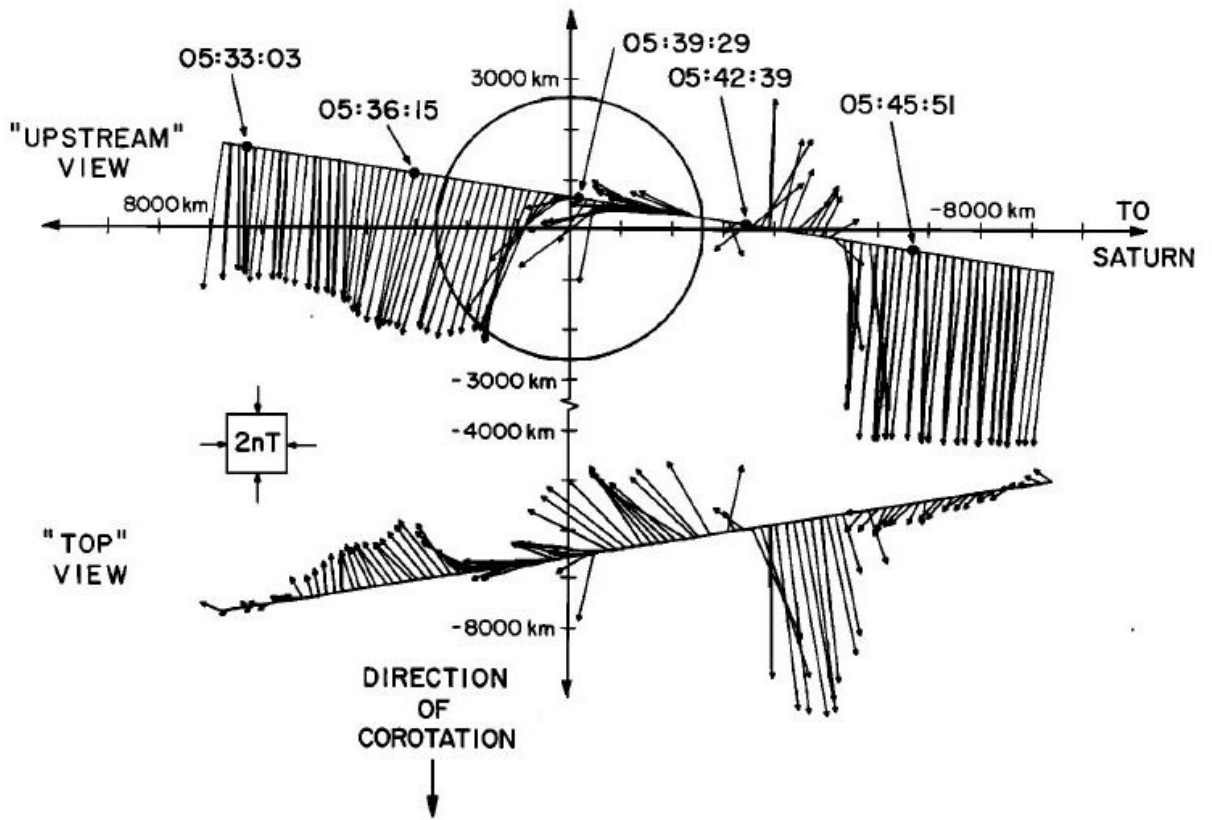

Figure 3: Plots of magnetic field vector components along the trajectory of *Voyager* -1 (*Ness et al.*, 1982)

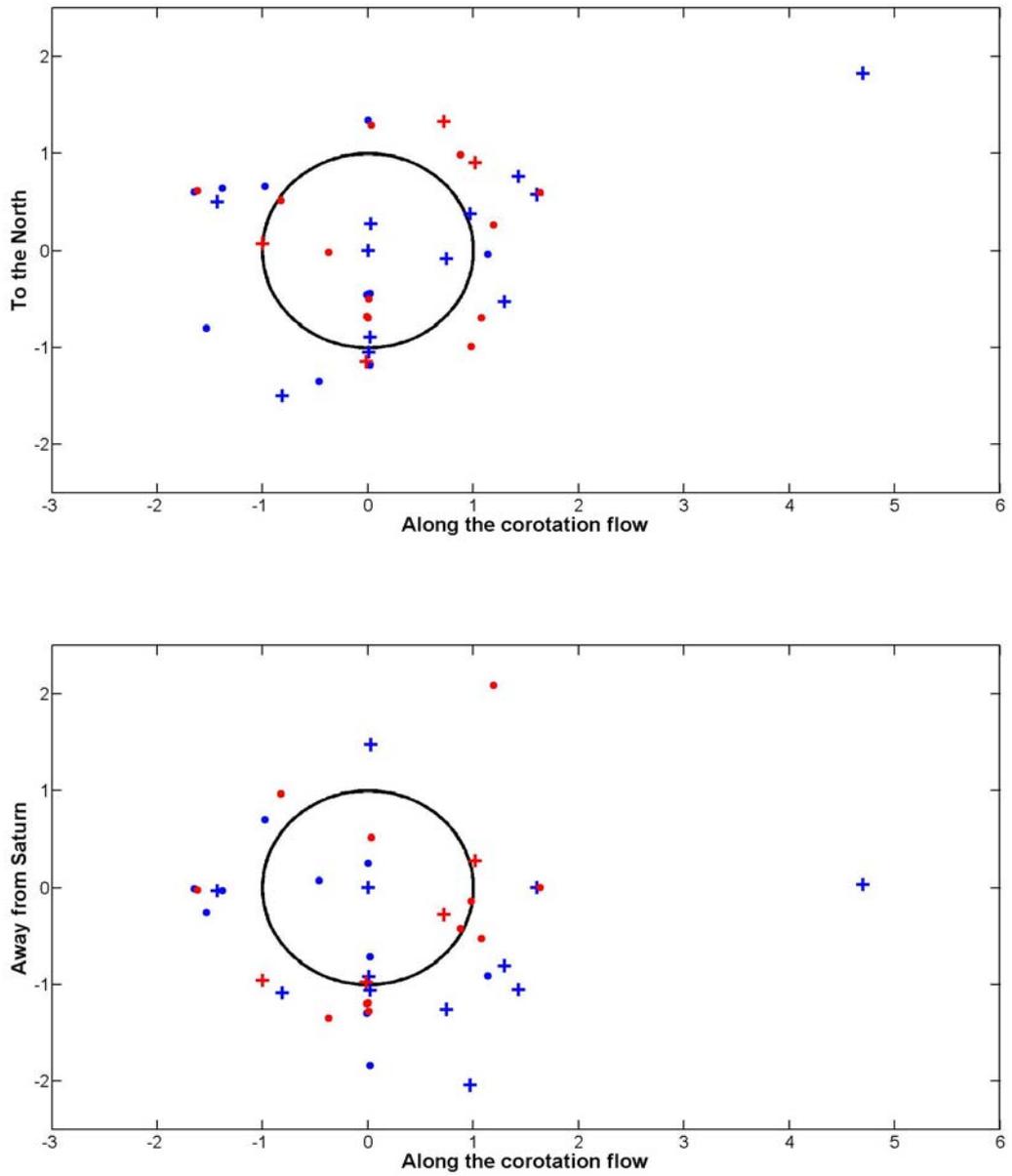

Figure 4: Positions of the Cassini closest approaches in Titan-centric coordinates. Circles (crosses) denote cases with small (moderate) perturbations in the corotating flow. Blue color corresponds to draping geometry and red color - to the violation of draping criterion.

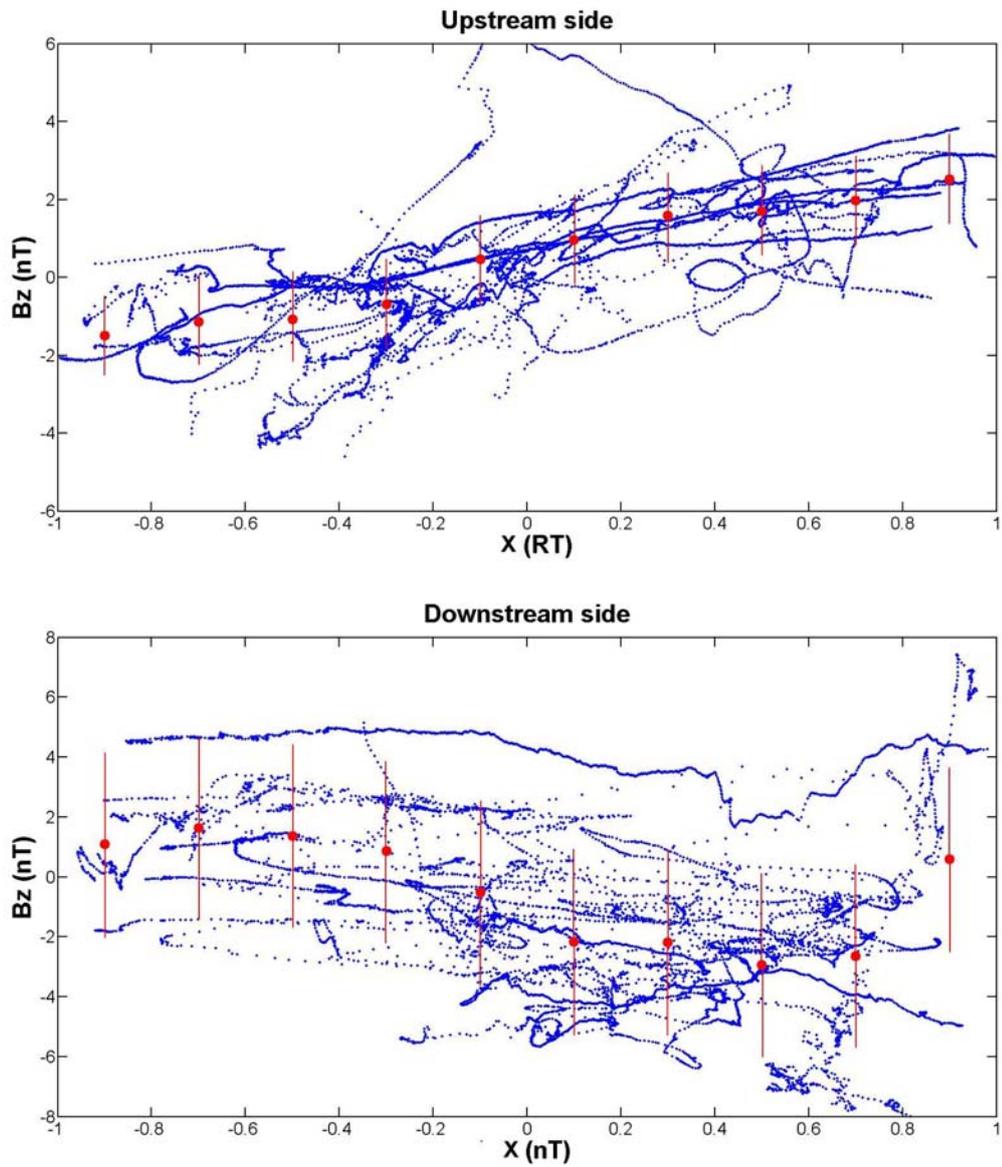

Figure 5: Dependence of the flow aligned magnetic field component $B_z$ on the distance along the $X_M$ axis. (Top) for the measurements upstream from the obstacle, (bottom) for downstream measurements. Blue dots - individual measurements, red dots - averages for 0.2 $R_T$ bins along the $X_M$ coordinate.

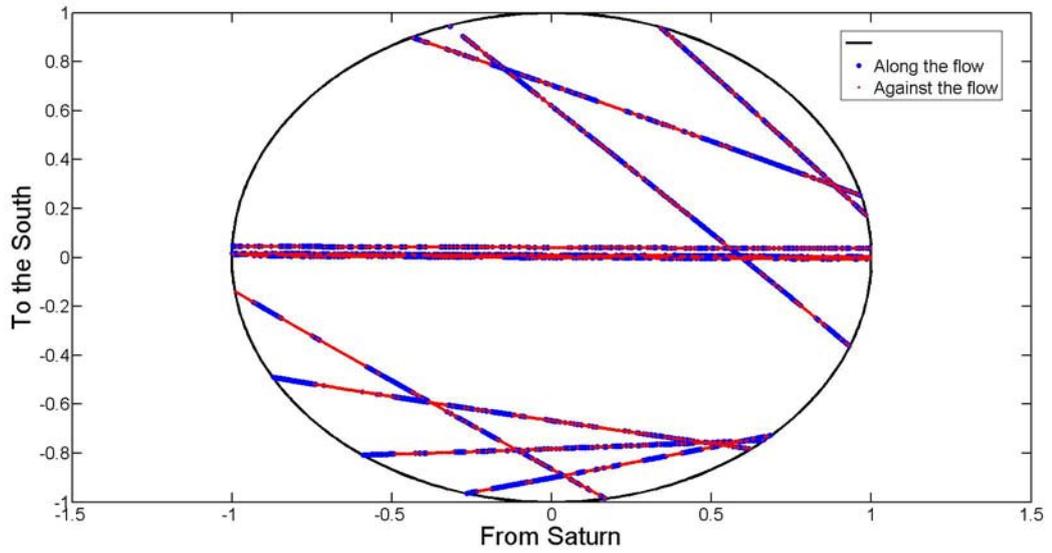

Figure 6: Direction of magnetic tension force in the Titan tail. (Blue dots along the flow, red dots against the flow, correspondingly).

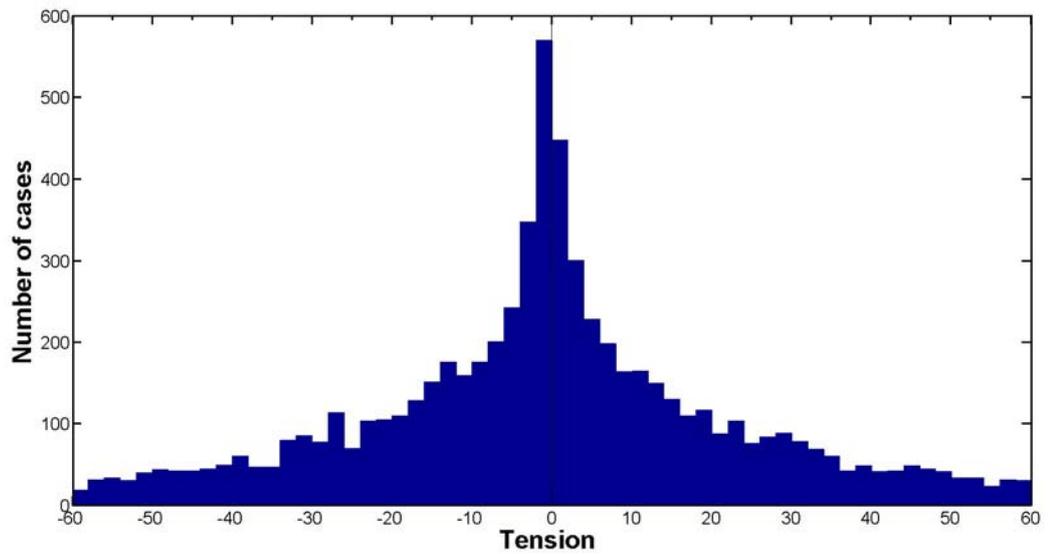

Figure 7: Histogram of distribution of magnetic tension in the region of Titan wake. Only the points at the distances less than 1 $R_T$ from the tail axis are taken into account.